\documentclass[10pt,conference]{IEEEtran}

\IEEEoverridecommandlockouts
\usepackage[table,xcdraw]{xcolor}
\usepackage{cite}
\usepackage{comment}
\usepackage[colorlinks=false,hidelinks=true]{hyperref}
\usepackage{balance}
\usepackage{amsmath,amssymb,amsfonts}
\usepackage{algorithmic}
\usepackage{graphicx}
\usepackage{textcomp}
\usepackage{xcolor}
\usepackage{xurl}
\usepackage{booktabs}
\usepackage{multirow}
\usepackage{diagbox}
\usepackage{setspace}
\usepackage{tcolorbox}
\usepackage{amsmath}
\usepackage{colortbl}
\usepackage{tablefootnote}
\tcbset{
	colback=black!5!white, colframe=black,
	notitle, 
	width={\linewidth+1pt},
	top=0pt,
	left=1pt,
	right=1pt,
	bottom=0pt,
	toprule=0.5pt,
	titlerule=1pt,
	bottomrule=0.5pt,
	leftrule=0.5pt,
	rightrule=0.5pt,
	after skip=6pt,
}

\begin{document}

\title{Is It Enough to Recommend Tasks to Newcomers? \\
Understanding Mentoring on Good First Issues}

%\thanks{Identify applicable funding agency here. If none, delete this.}
%}
%\author{ANONYMOUS}
\author{\IEEEauthorblockN{Xin Tan}
\IEEEauthorblockA{\textit{School of Computer Science and Engineering} \\
\textit{Beihang University}\\
\textit{State Key Laboratory of Software Development Environment}\\
Beijing, China \\
xintan@buaa.edu.cn}
\and
\IEEEauthorblockN{Yiran Chen}
\IEEEauthorblockA{\textit{ShenYuan Honors College} \\
\textit{Beihang University}\\
Beijing, China \\
aurora\_cccyr@buaa.edu.cn}
\and
\IEEEauthorblockN{Haohua Wu}
\IEEEauthorblockA{\textit{ShenYuan Honors College} \\
\textit{Beihang University}\\
Beijing, China\\
oliverhaohuawu@buaa.edu.cn}
\and
\IEEEauthorblockN{Minghui Zhou}
\IEEEauthorblockA{\textit{School of Computer Science, Peking University} \\
\textit{Key Laboratory of High Confidence Software Technologies}\\
\textit{Ministry of Education}\\
Beijing, China\\
zhmh@pku.edu.cn}
\and
\IEEEauthorblockN{Li Zhang}
\IEEEauthorblockA{\textit{School of Computer Science and Engineering} \\
\textit{Beihang Univeristy}\\
\textit{State Key Laboratory of Software Development Environment}\\
Beijing, China \\
Corresponding author: lily@buaa.edu.cn}
}
\maketitle

\begin{abstract}
Newcomers are critical for the success and continuity of open source software (OSS) projects. To attract newcomers and facilitate their onboarding, many OSS projects recommend tasks for newcomers, such as good first issues (GFIs). Previous studies have preliminarily investigated the effects of GFIs and techniques to identify suitable GFIs. However, it is still unclear whether just recommending tasks is enough and how significant mentoring is for newcomers. To better understand mentoring in OSS communities, we analyze the resolution process of 48,402 GFIs from 964 repositories through a mix-method approach. We investigate the extent, the mentorship structures, the discussed topics, and the relevance of expert involvement.
We find that $\sim$70\% of GFIs have expert participation, with each GFI usually having one expert who makes two comments. Half of GFIs will receive their first expert comment within 8.5 hours after a newcomer comment. Through analysis of the collaboration networks of newcomers and experts, we observe that community mentorship presents four types of structure: centralized mentoring, decentralized mentoring, collaborative mentoring, and distributed mentoring. As for discussed topics, we identify 14 newcomer challenges and 18 expert mentoring content. By fitting the generalized linear models, we find that expert involvement positively correlates with newcomers' successful contributions but negatively correlates with newcomers' retention. Our study manifests the status and significance of mentoring in the OSS projects, which provides rich practical implications for optimizing the mentoring process and helping newcomers contribute smoothly and successfully. 
\end{abstract}


\section{Introduction}
Open source software (OSS) projects are playing an increasingly critical role in the software industry. Different from traditional software development, OSS development serves as a prominent example of open collaboration, where many developers around the world work together online to develop a software product~\cite{raymond1999cathedral}. While OSS development makes full use of group wisdom, its loose organizational structure causes a high contributor turnover rate~\cite{zhou2012make}. Therefore, OSS projects rely on a continuous influx of newcomers to ensure their sustainable and healthy development~\cite{zhou2012make}. Meanwhile, newcomers are also eager to participate in OSS development because it is a fantastic way to learn real-world software development skills~\cite{ye2003toward}. However, newcomers face many barriers when making their first contributions to OSS projects, leading in many cases to dropouts~\cite{mendez2018open}.

To support newcomers' onboarding, OSS communities have put forward various strategies, e.g., providing readme files and contributing guidelines~\cite{kobayakawa2017github}. Although these strategies and mechanisms can indeed reduce the barriers to newcomers' onboarding, newcomers still have difficulty in making their first contributions due to their limited experience or high technical threshold~\cite{Steinmacher2014Preliminary}. To help newcomers contribute, GitHub recommends OSS communities to label ``good first issues'' (GFIs) in the repository, thus highlighting opportunities for people to contribute.\footnote{\url{https://docs.github.com/en/communities/setting-up-your-project-for-healthy-contributions/encouraging-helpful-contributions-to-your-project-with-labels}} Since 2009, there are more and more projects labeling GFIs, especially for extremely popular projects~\cite{tan2020first}. In recent years, due to the wide application of machine learning and deep learning technologies, some automatic classification algorithms even have been proposed in order to help maintainers label GFIs~\cite{Xiao2022GoodFirstIssues, huang2021characterizing}.\footnote{\url{https://github.blog/2020-01-22-how-we-built-good-first-issues/}} Thus it can be seen that OSS communities and researchers have spent a lot of effort on task recommendations for newcomers. However, is it enough to recommend tasks to newcomers? Do they need mentoring from experts? If the answer is yes, how to mentor newcomers? These are critical questions but are still unclear, which may not only make the task recommendation less effective but also hinder newcomers' onboarding.

Previous studies on OSS mentoring mainly focus on open source programs, e.g., Google summer of code (GSoC)~\cite{trainer2017mentoring, balali2018newcomers,trainer2013big}. In these programs, students or people who are subject to systemic bias will receive the guidance of mentors to learn real-world software development and other domain experience. However, research finds that in OSS projects, formal mentoring for newcomers is rare because mentoring newcomers takes plenty of time and energy while experts are usually extremely busy~\cite{ducheneaut2005socialization}. Besides, most of these studies are based on interview or questionnaire data, with little analysis of historical data on the mentoring process. This results in a lack of objective knowledge of the daily mentoring behavior in OSS communities. To bridge this gap, we investigate OSS mentoring based on the analysis of the GFIs resolution process, and put forward the following research questions: 

\textbf{RQ1 (Expert Involvement)}: \textit{To what extent do experts involve in the resolution of GFIs?}

\noindent Based on the quantitative analysis of the resolution process of 48,402 GFIs, we find that $\sim$70\% of GFIs have expert participation. Each GFI usually has one expert and one newcomer participants, and each person %only leaves
often makes two comments. For half of GFIs, the expert response is received within 8.5 hours.

\textbf{RQ2 (Mentorship Structure)}: \textit{What structures does mentorship take in OSS communities?}
%\textbf{RQ2 (Mentorship Type)}: \textit{What are the mentorship types staged in OSS communities?}

\noindent We draw on bipartite graph to build the mentorship networks of OSS communities and find four types, i.e., centralized/ decentralized/ collaborative/ distributed mentoring. Among them, collaborative mentoring accounts for more than one-third. It indicates that these communities have several experts collaborating when helping newcomers.

\textbf{RQ3 (Discussed Topic)}: \textit{What topics are often discussed during the mentoring?}

\noindent We %qualitatively analyze 
employ thematic analysis method to analyze
43 highly discussed GFIs, including a total of 1,065 comments from 131 newcomers and 98 experts. Combining with an online survey, we identify 32 topics, belonging to two themes: newcomer challenges and expert mentoring.

%\textbf{RQ4 (Importance)}: \textit{What are the relationships between experts' participation and newcomers' onboarding?}
\textbf{RQ4 (Relevance)}: \textit{Is experts' participation associated with newcomers' onboarding?}

\noindent We build the generalized linear models to explore the relevance of experts' participation. We find that experts' participation has a positive relationship with newcomers' successful contributions. However, it is negatively correlated with newcomers' retention, instead, the results imply that newcomers' ability and willingness are the main factors.

The major contributions of this paper are as follows:

\begin{itemize}
\item To the best of our knowledge, this is the first comprehensive study on the daily mentoring behavior for newcomers' onboarding in OSS communities. 
\item We provide practical insights on mentoring in OSS communities and shed light on directions for future research.
\end{itemize}

%We organize the remainder of the paper as follows. Section~\ref{related_work} presents related work; Section~\ref{dataset} introduces our dataset. Section~\ref{RQ1} to Section~\ref{RQ4} present methods and answers to each of the four research questions. Section~\ref{implications} discusses the implications of findings. Section~\ref{threats} presents threats to validity and Section~\ref{conclusion} concludes the paper.

\section{Related Work}\label{related_work}
\subsection{Barriers to Newcomers' Onboarding}
Substantial work investigates the barriers to newcomers' onboarding. Researchers find that these barriers involve two categories: technical barriers and non-technical barriers~\cite{balali2018newcomers,azanza2021onboarding}. Lack of the necessary domain knowledge and programming skills are the main technical barriers faced by newcomers in their initial activities~\cite{lee2017understanding,shibuya2009understanding}. Some non-technical barriers also hinder newcomers’ first contributions, e.g., lack of communication skills and interpersonal skill~\cite{tan2019communicate,ducheneaut2005socialization}. Through a systematic literature review, Steinmacher et al.~\cite{steinmacher2015systematic} identified 15 different barriers of newcomers grouped into five categories: social interaction, newcomers’ previous knowledge, finding a way to start, documentation, and technical hurdles. Some studies focus on the barriers faced by the newcomers who are subject to systematic bias, among which gender bias is a critical issue in such male-dominated communities~\cite{QiuGoing2019}. For example, through a field study with five teams of software professionals, Padala et al.~\cite{padala2020gender} find that gender biases exist in 73\% of all the newcomer barriers the professionals identified.

The above studies indicate that newcomers' onboarding is fraught with difficulties. Our study complements these works by investigating the barriers/difficulties that newcomers may face when completing their initial contributions with the help of experts. It is important to note that as our results are derived from historical data, they are more fine-grained and therefore more likely to have practical implications.

\subsection{Supporting Newcomers' Onboarding}
To support newcomers' onboarding, some theories and strategies have been proposed. Through modeling newcomers' initial behavior and interaction, researchers try to reveal the factors related to newcomers' long-term contribution~\cite{zhou2014will, zhou2012make, bao2019large, jensen2011joining}. For example, Jensen et al.~\cite{jensen2011joining} find that receiving timely responses, especially within 48 hours, is positively correlated with newcomers' future participation. Based on the interview and analysis of the prior work, Steinmacher et al.~\cite{steinmacher2018let} propose guidelines assisting newcomers onboard including contribution process guidelines, social behavior guidelines, and technical guidelines. Task recommendation for newcomers is also a hot issue focused on by researchers recently~\cite{tan2020first, Xiao2022GoodFirstIssues, huang2021characterizing}.

Some researchers concentrate on online mentoring for newcomers. For example, Panichella~\cite{panichella2015supporting} investigates how to recommend mentors for newcomers. Summer of code programs (i.e., GSoC) is a kind of typical case for researchers to study the mentoring process of newcomers~\cite{trainer2017mentoring, balali2018newcomers,trainer2013big}. However, most of these studies are based on surveys and interviews, lacking an objective analysis of mentoring historical data. A more serious concern is that formal mentoring is uncommon in OSS communities~\cite{ducheneaut2005socialization, feng2022implicit}. Different from these studies, we make a comprehensive understanding of mentoring in GFI resolution process, which is necessary for understanding daily mentoring behavior in OSS communities and further helping newcomers to join OSS projects.

\section{Dataset}\label{dataset}
\begin{table}[]
\centering
\scriptsize
\caption{Project Statistics in Constructed Dataset}
%%\vspace{-0.2cm}
\begin{tabular}{crrrrr}
\toprule
& \multicolumn{1}{c}{\multirow{2}{*}{\begin{tabular}[c]{@{}c@{}}Age\\ (months)\end{tabular}}} & \multicolumn{1}{c}{\multirow{2}{*}{\begin{tabular}[c]{@{}c@{}}Contributors\\ (authors)\end{tabular}}} & \multicolumn{1}{c}{\multirow{2}{*}{Commits}} & \multicolumn{1}{c}{\multirow{2}{*}{Issues}} & \multicolumn{1}{c}{\multirow{2}{*}{\begin{tabular}[c]{@{}c@{}}GFIs\\ (closed)\end{tabular}}} \\
& \multicolumn{1}{c}{} & \multicolumn{1}{c}{} & \multicolumn{1}{c}{} & \multicolumn{1}{c}{} & \multicolumn{1}{c}{} \\
\midrule
Total &  44,050 & 157,792 & 7,132,566 & 3,710,986 & 48,402 \\
Mean & 46.76 & 166.62 & 7,531.75 & 3,918.68 & 51.11 \\
Med. & 39 & 59 & 2,687 & 1,346 & 34 \\
Min. & 0 & 1 & 2 & 36 & 1 \\
Max. & 148 & 11,168 & 217,109 & 108,538 & 1,504 \\
Std. & 32.58 & 471.02 & 15,105.84 & 8,510.83 & 72.59 \\
\bottomrule
%\vspace{-0.2cm}
\end{tabular}
\label{tab:Project_Statistics}
\end{table}

We conduct this study based on the GHTorrent database that provides an offline mirror of GitHub data~\cite{gousios2013ghtorent}. At the time of data collection, its latest version is ``2021-03-06". Therefore, we use this version to construct our dataset. Among all the 189,524,128 repositories, we find that 96,321 non-forked repositories have ever reported GFIs. However, most repositories only reported a small number of GFIs: median: 1 GFI. %maximum: 1780 GFIs. 
To select representative OSS projects that adopt the GFI mechanism, we choose the top 1\% repositories with most GFIs (i.e., $\#GFIs\geqslant30$) as our initial dataset, including 964 repositories and 68,652 GFIs. Then, we filter the GFIs that are still in progress, i.e., only keep the closed GFIs. We also extract the comments of the GFIs. Because developers may leave comments on both issues and associated solutions, we write a script that can not only extract the comments under GFIs but also the comments under the related pull requests (PR) or commits through the hyperlinks contained in the issues' comments. Eventually, we obtain a dataset containing 947 repositories, 48,402 closed GFIs, and 286,371 comments. The comments contain 176,674 issue comments and 109,697 PRorCommit comments\footnote{Considering both PR comments and commit comments can be treated as solution-oriented, we do not distinguish them in this paper, instead denote them as PRorCommit comments.}. Table~\ref{tab:Project_Statistics} shows the basic information of the projects in the constructed dataset.\footnote{We use ``repository'' and ``project'' interchangeably in the paper.} We can see that most projects are in large scale but the GFIs take a small proportion among all the issues, which reflects the necessity of investigating how to make good use of such precious issues to help newcomers' onboarding.

\begin{figure}[htp]
\centering 
\includegraphics[width = 7.5cm]{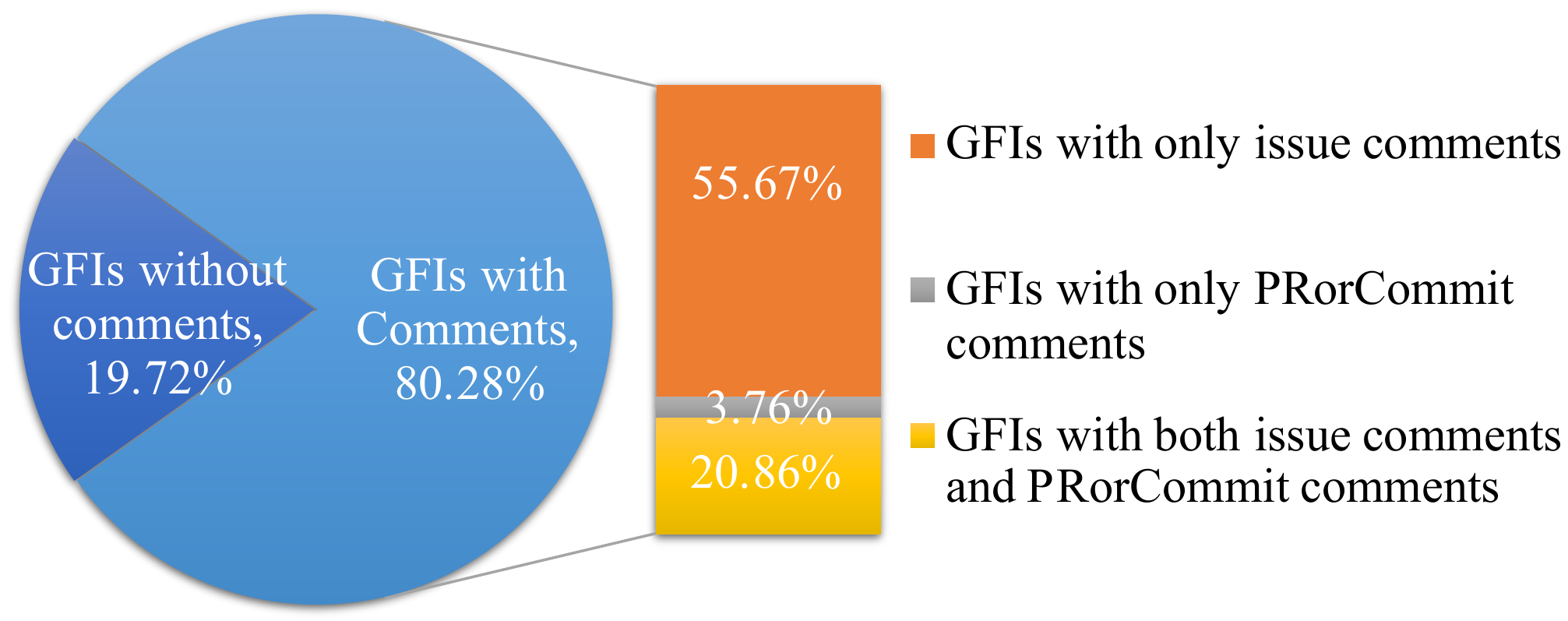}
%\vspace{-0.2cm}
\caption{Proportion of GFIs with/without Comments}
\label{fig:comments_distribution}
\end{figure}

Fig.~\ref{fig:comments_distribution} shows the distribution of the GFIs' comments. There are more than 80\% of GFIs have comments, among them 55.67\% only have issue comments and 20.86\% have both issue comments and PRorCommit comments. These 286,371 comments serve as the primary data source for understanding mentoring on GFIs. 

\section{RQ1: Expert Involvement}\label{RQ1}
We aim to understand to what extent experts involve in the resolution of GFIs. %All the analyses are quantitative.

\subsection{Methodology}
We answer this question from three perspectives: 1) who participate in solving GFIs; 2) how much effort experts spend in solving GFIs, and 3) how quickly experts respond. 
To this end, we need to determine the identities of the participants in the GFI resolution process. Referring to the related studies, we give the following definitions, and utilize them throughout the study. These definitions are for a single project.
\begin{itemize}
    \item \textbf{Experts}: a developer was among the top 20\% of contributors (considering \#commits authored) in the year before she/he made the comment~\cite{mockus2002two}.
    \item \textbf{Newcomers}: the number of commits that a developer contributed to the project is less than three~\cite{tan2020first}.
    \item \textbf{General Developers}: the developers who are neither experts nor newcomers.
\end{itemize}

\subsection{Results}
\subsubsection{Participants}
\begin{figure}[htp]
\centering 
\includegraphics[width = 6cm]{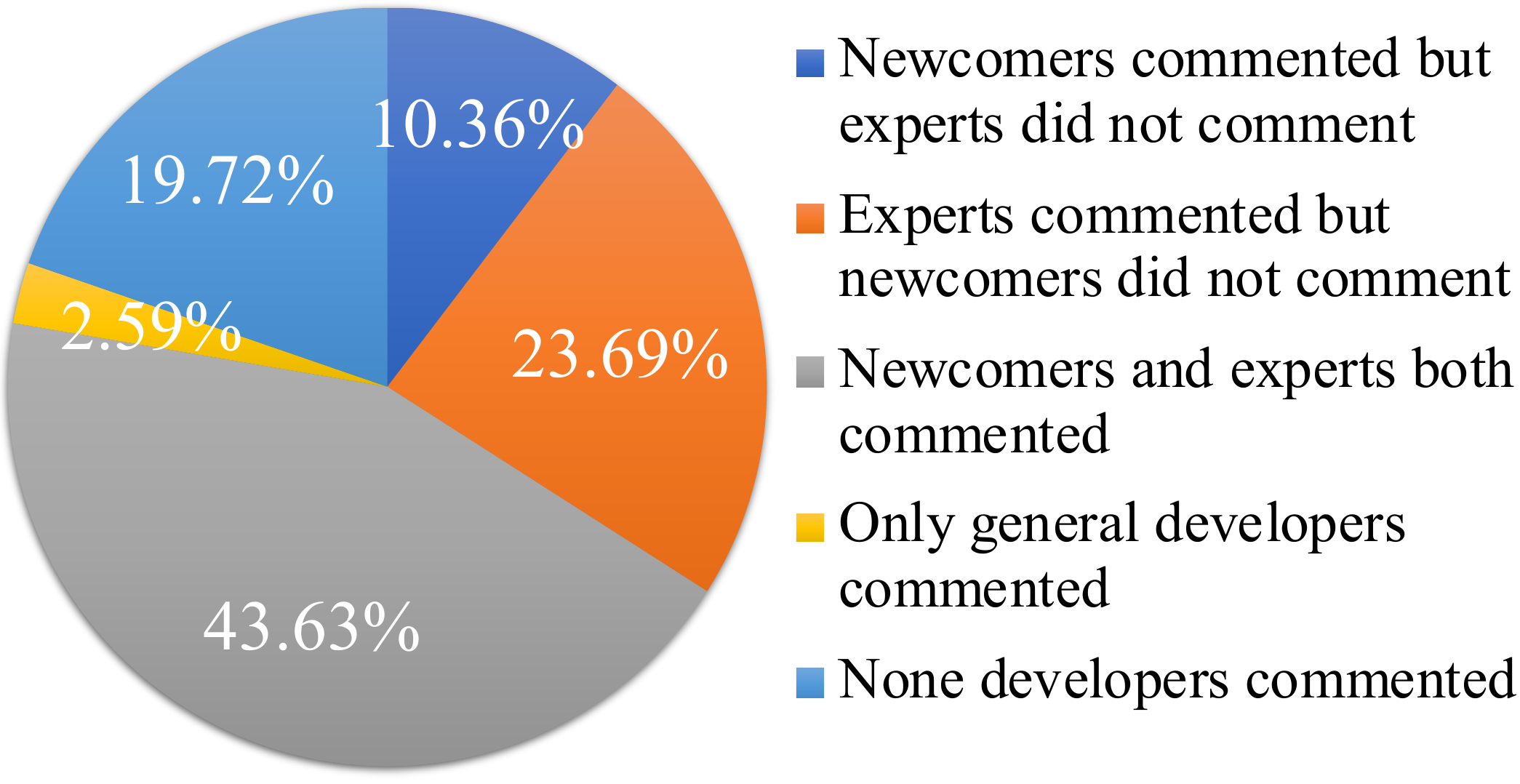}
%\vspace{-0.2cm}
\caption{Proportion of GFIs with Different Participants} %be clear about the data points: #GFIs?
\label{fig:RQ1_participants}
\end{figure}
As shown in Fig.~\ref{fig:RQ1_participants}, during the resolution process, 43.63\% of GFIs receive comments from both newcomers and experts, whereas 10.36\% of GFIs only receive newcomers' comments. Few (only 2.59\%) GFIs receive comments only from general developers. In total, nearly 70\% of GFIs have expert participation, which indicates that experts usually participate in the GFI resolution process.

\subsubsection{Efforts} 
\begin{figure}[htp]
\centering 
\includegraphics[width = 6.5cm]{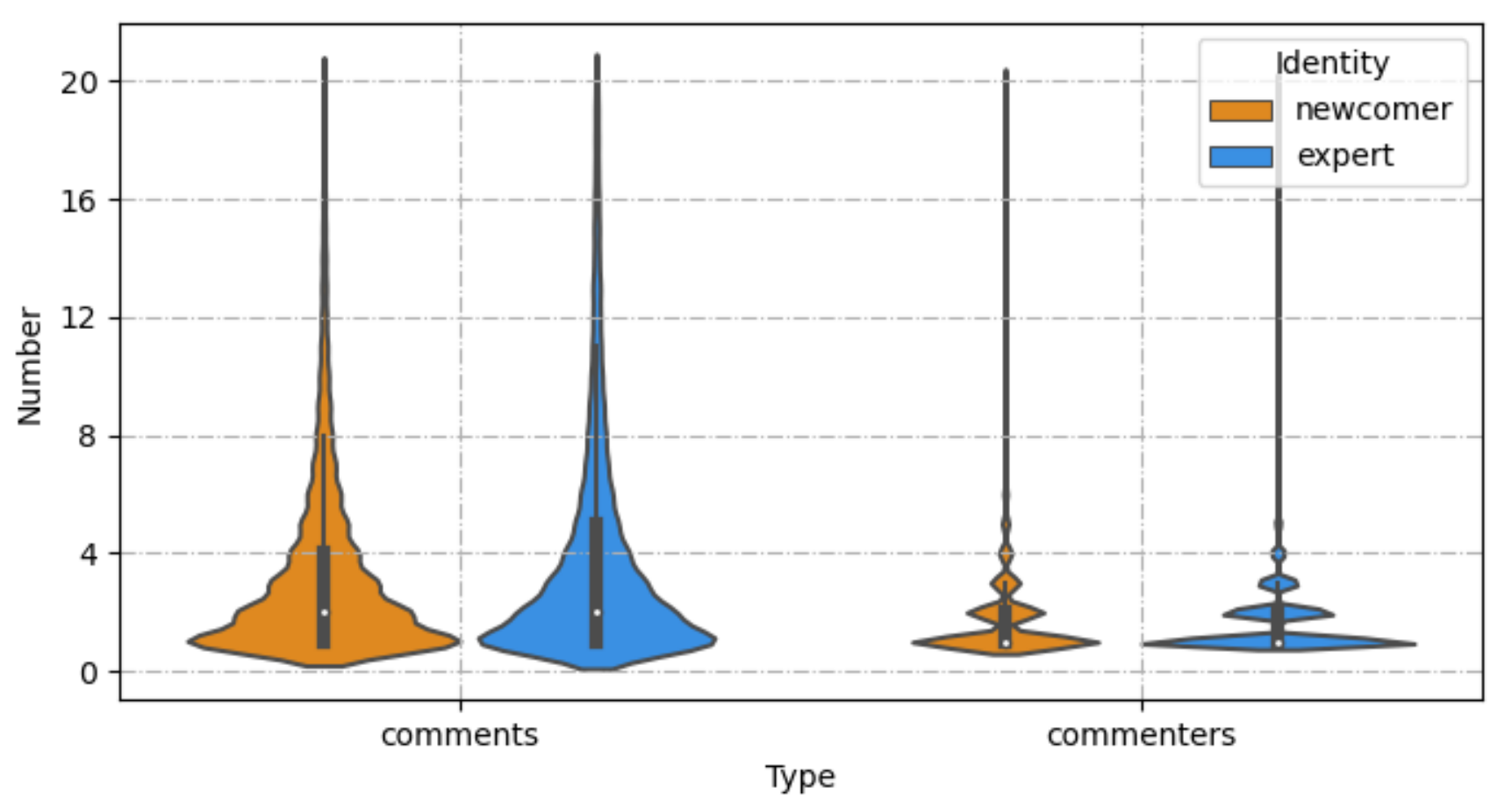}
%\vspace{-0.2cm}
\caption{Distribution of Number of Newcomers and Experts Involved in GFI Resolution Process and Their Comments (Remove Zero Values and Outliers)} %each data point is a GFI?
\label{fig:RQ1_effort}
\end{figure}
We define the efforts of a developer as the number of comments posted under the GFI. As shown in Fig.~\ref{fig:RQ1_effort}, %in the process of solving GFIs, 
each GFI %(with the participation of newcomers or experts) %%what does this "with" mean?
usually has one expert and one newcomer %who 
participated, 
and each person %only 
posts two comments. A few GFIs involve many newcomers and experts with many comments. We also observe that the amount of effort newcomers and experts put into solving GFIs is basically equal.

\subsubsection{Response Time} 
\begin{comment}
\begin{figure}[htp]
\centering 
\includegraphics[width = 5cm]{pic/response_time.pdf}
\caption{Distribution of Expert Response Time (Remove Outliers)}
\label{fig:response_time}
\end{figure}
\end{comment}
We define response time as the time interval between newcomers' first comment time $t_N$ and experts' first comment time $t_E$, where $t_E > t_N$. %As shown in Fig.~\ref{fig:response_time}, 
Among the GFIs with expert participation, we find that a quarter of GFIs receive experts' comments within 41.25 minutes after newcomers commented. As for half %of GFIs 
and 75\% of GFIs, this value becomes 8.50 hours and 4.25 days respectively. However, 10\% of GFIs receive an expert's comment more than 56 days later. We notice that most of these cases are because the newcomers do not post clear questions, or the issues are stale due to various reasons, e.g., new versions, or the experts are too busy to respond until the robot alerts them.
%which may seriously reduce the enthusiasm of newcomers and even make the response ineffective.
%%it might be interesting to explore what makes the delay: is it because the newcomer comment too naive or because the expert being busy or lazy?

\vspace{4pt}
\noindent\begin{tcolorbox}
\noindent\emph{\textbf{Summary for RQ1:}}
Expert participation is common in GFIs' resolution process: 70\% of GFIs have expert involvement but their effort is limited. By median, each GFI usually has one expert participating, and this expert typically makes two comments. The expert posts the first comment usually within 8.5 hours after the newcomer's first comment. The response time for a few GFIs is more than two months. %%what makes the delay? it might be irrelevant
\end{tcolorbox}

\section{RQ2: Mentorship Structure}\label{RQ2}
This question aims to understand the social networks of mentorship in OSS communities. \textcolor{black}{We want to know what structures of mentorship are presented, such as multiple experts or single experts, and whether there are relationships between different mentorship structures and characteristics of OSS communities. Answering this question can enrich our understanding of mentorship in OSS communities.}
\subsection{Methodology}
Social networks focus on relationships among social entities and are important to investigate the patterns and implications of these relationships.~\cite{wasserman1994social} To understand the mentorship in OSS communities, we draw on the social network theory to define the mentorship network of an OSS community as an integer weighted bipartite graph~\cite{asratian1998bipartite}: 
\begin{equation}
\begin{array}{l}
    G = \{U \sqcup V,E, w_e, w_n\}\\
    U = \{u_1, u_2,...,u_{n_1}\} \\
    V = \{v_1, v_2,...,v_{n_2}\} \\
    E = \{<u_j, v_k>\} \\
    w_n:U,V\rightarrow \mathbb{Z}^{+};\ w_e:E\rightarrow \mathbb{Z}^{+}
\end{array}
\end{equation}
where $u_j$ is a newcomer, $v_k$ is an expert, $<u_j, v_k>$ means that $u_j$ and $v_k$ posted comments on the same GFI, the weight of node $u_j$ or $v_k$ represents \#GFIs that $u_j$ or $v_k$ commented on, and the weight of the edge $<u_j, v_k>$ represents \#GFIs that $u_j$ and $v_k$ both commented on. 

To explore the patterns of mentorship structure, we randomly sampled 275 repositories from 964 repositories (confidence level: 95\%; margin of error: 5\%) and use NetworkX\footnote{\url{https://networkx.org/}} to visualize their networks. Then, we performed an open picture sorting, a kind of open card sorting technique on image data~\cite{lobinger2019picture,spencer2009card}. Firstly, we created one picture card for each repository network. %Then, the second and third authors separately grouped these cards into potential categories considering their structure commonalities. The overall Kappa valuebetween the two labelers is 0.77, which indicates substantial agreement~\cite{warrens2015five}. After the labeling process, the first three authors discussed together to solve the disagreement and define the final names and descriptions for these categories. 
Then, the first three authors grouped these cards to create categories of pictures with structure commonalities and to create names and descriptions for these categories. During this process, the first three authors discussed single items together in order to eventually agree on the categories built and on their meanings.

To understand the topology of different mentorship structures, we calculate the following graph metrics. Except for the first metric, we calculate the average metric value of all nodes in the graph.
\begin{itemize}
    \item $ratio\_of\_expert$: proportion of experts in mentorship;
    \item $weighted\_degree$: sum of weights of connected edges;
    \item $degree\_centrality$~\cite{scott2011sage}: the fraction of developers a certain developer is connected to;
    \item $closeness\_centrality$~\cite{scott2011sage}: reciprocal of the sum of the shortest paths from a developer to other developers; 
    \item $betweenness\_centrality$~\cite{brandes2008variants}: the sum of the fraction of all-pairs shortest paths that pass through a developer.
\end{itemize}
We also calculate the project-related metrics for each network: \#authors, \#commits, \#issues, \#months (project age), and \#stars.

\subsection{Results}
\begin{figure*}[htp]
\centering 
\includegraphics[width = 16cm]{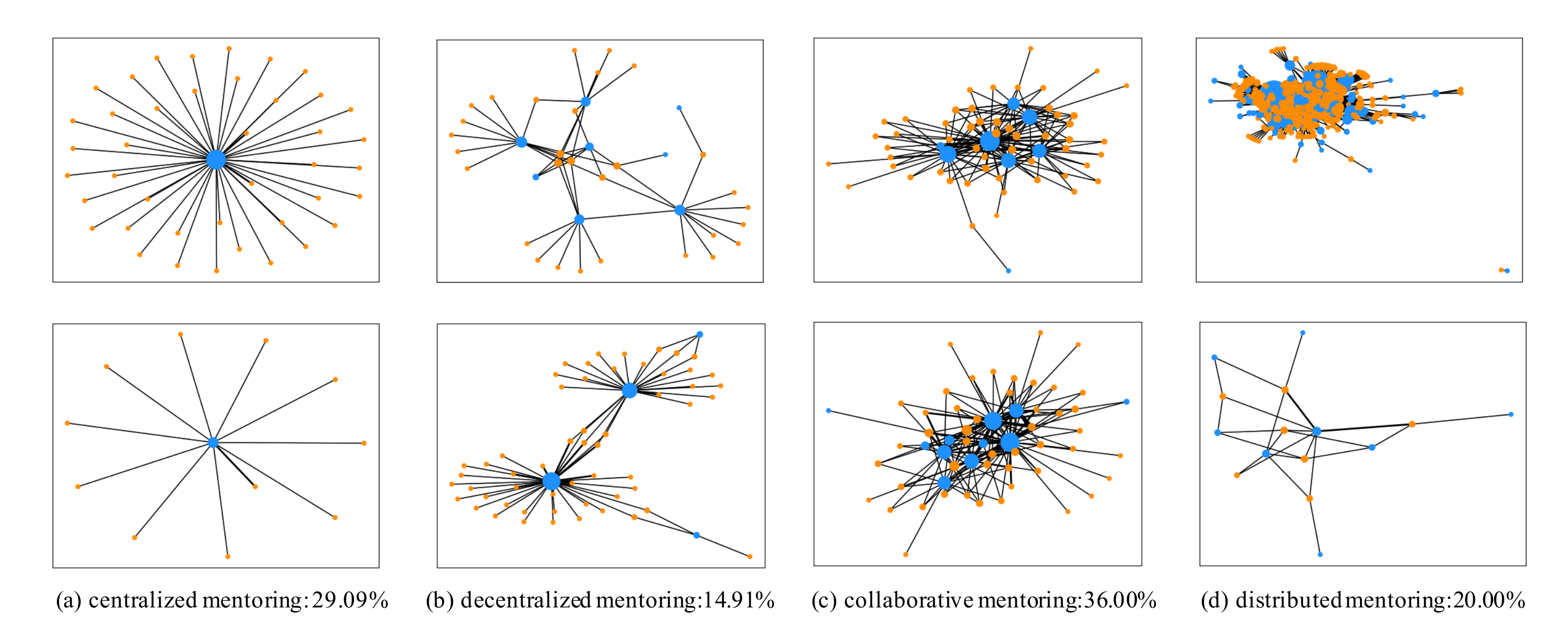}
%\vspace{-0.3cm}
\caption{Examples of Different Mentorship Structure and their Proportions (Blue Nodes: Experts, Orange Nodes: Newcomers)}
%\vspace{-0.4cm}
\label{fig:network_type}
\end{figure*}

As shown in Fig.~\ref{fig:network_type}, we identify four types of mentorship structures in OSS communities.

\textbf{Centralized Mentoring.} This is a type of highly centralized network~\cite{gavish1985augmented} where the community is built around a single, central expert node. The communities belonging to this type mainly have one expert who provides guidance for newcomers. \textcolor{black}{This may indicate that these communities are smaller or have a limited number of experts.}

\textbf{Decentralized Mentoring.} This is a type of decentralized network~\cite{dabek2004vivaldi} where the community has several relatively independent expert nodes connected with newcomer nodes. The communities belonging to this type generally have several experts who provide guidance for newcomers. Each expert in a community is relatively independent. \textcolor{black}{This may indicate that each expert has their own area of expertise or is responsible for a certain module.}

\textbf{Collaborative Mentoring.} This is a type of moderate centralized network where the community is organized by several relatively dependent expert nodes. The communities belonging to this type usually have several experts who collaborate to provide guidance for newcomers. \textcolor{black}{This may indicate that these communities have more versatile experts who are willing to collaborate to provide guidance in multiple areas.}

\textbf{Distributed Mentoring.} In this type of communities, all newcomer nodes are split evenly among many equal expert nodes across the entire network. The experts who provide guidance for newcomers are random. \textcolor{black}{The communities do not seem to set up dedicated experts in place to guide newcomers.}

Among the four types of mentorship structures, collaborative mentoring accounts for the highest proportion (36.00\%). It means that more than one-third of communities have several experts collaborating with each other when helping newcomers. Nearly 30\% of the communities belong to centralized mentoring, i.e., only one expert is mainly responsible for interaction with newcomers. The type that accounts for the least is decentralized mentoring. It means that fewer communities (14.91\%) are able to form more independent mentoring groups, which may be related to the modularity of projects.

\begin{figure}[htp]
\centering 
\includegraphics[width = 7cm]{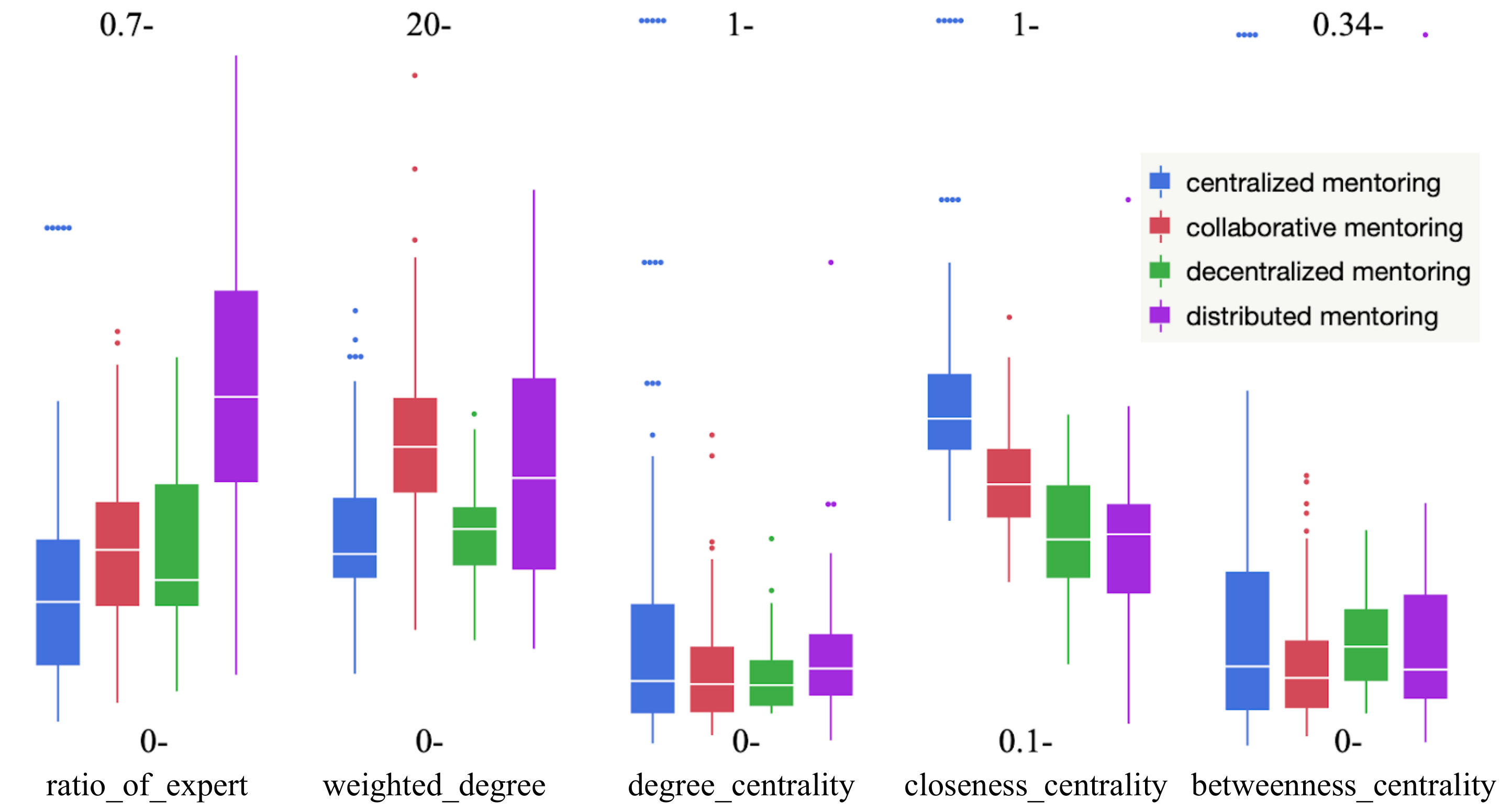}
%\vspace{-0.3cm}
\caption{Topology Characteristics of Different Mentorship Structures}
\label{fig:graph_metrics}
\end{figure}

Fig.~\ref{fig:graph_metrics} shows the \textbf{graph metrics} distributions of four types of mentorship structures. We can see that the communities belonging to distributed mentoring have a higher $ratio\_of\_experts$, whereas the communities belonging to centralized mentoring have much lower values. This is natural because, in distributed mentoring communities, many experts interact with newcomers, whereas in centralized mentoring communities, there is usually only one expert who helps newcomers. For the $weighted\_degree$,  the communities belonging to collaborative mentoring and distributed mentoring have higher values than the communities belonging to centralized mentoring and decentralized mentoring. It is  because that in centralized mentoring and decentralized mentoring, each newcomer usually has a mentoring relationship with only one expert, whereas in collaborative mentoring and distributed mentoring communities, multiple experts are usually involved during a mentoring process. For $closeness\_centrality$, it is obvious that the communities belonging to centralized mentoring have higher values because the main expert makes all the contributors closer. For $betweenness\_centrality$, we can see that communities belonging to decentralized mentoring have higher values because several independent experts serve as bridges from one part of a network to another. As for $degree\_centrality$, the four types of networks are similar.

\begin{figure}[htp]
\centering 
\includegraphics[width = 7cm]{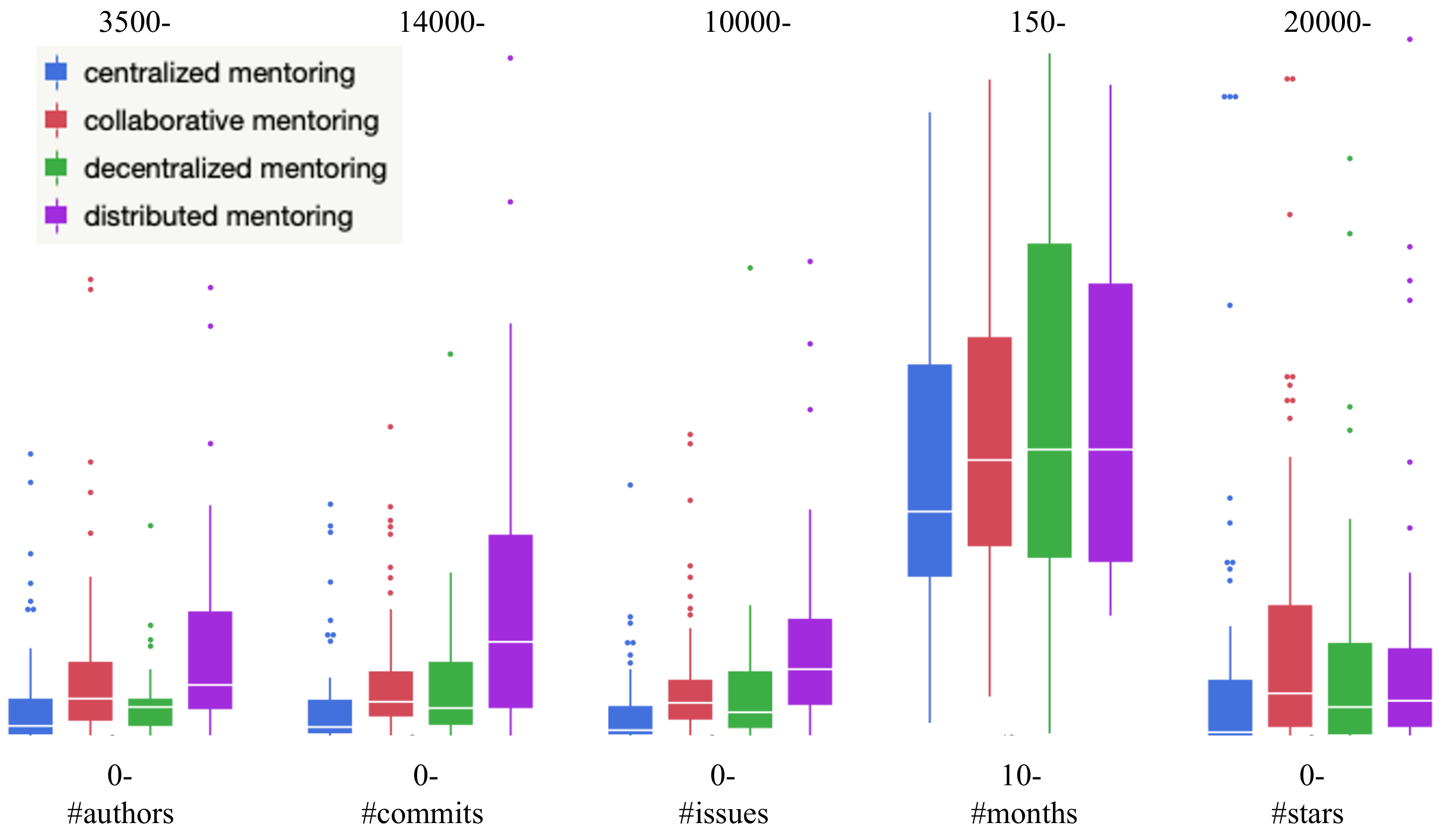}
%\vspace{-0.3cm}
\caption{Project Characteristics of Different Mentorship Structures}
%\vspace{-0.3cm}
\label{fig:project_metrics}
\end{figure}

Fig.~\ref{fig:project_metrics} shows the \textbf{project metrics} distributions of four types of mentorship structures. We can see that the communities belonging to centralized mentoring are usually smaller (\#authors, \#commits, \#issues), younger, and less popular. For these communities, there is usually only one core expert responsible for mentoring.
On the contrary, the communities belonging to distributed mentoring are bigger and more popular, so there are many experts helping newcomers.
It is worth noting that the mentorship structures of more popular projects are more likely to be collaborative mentoring, meaning that there are multiple experts collaborating to provide guidance for newcomers.

\vspace{4pt}
\noindent\begin{tcolorbox}
\noindent\emph{\textbf{Summary for RQ2:}}
There are four types of mentorship structures in OSS communities.
Collaborative mentoring is the most common type, followed by centralized mentoring, distributed mentoring, and decentralized mentoring. These four types differ in both topology and project characteristics. 
\end{tcolorbox}

\section{RQ3: Discussed Topics}\label{RQ3}
This question aims to extract the topics discussed by experts and newcomers during the GFI resolution process.
\subsection{Methodology}
\textcolor{black}{Our method has two stages. In the first stage, the goal is to obtain topics as rich as possible without care of their frequencies. Therefore, we focus on the ``highly discussed'' GFIs.}
We refer to the study of Jason et A.~\cite{tsay2014let} to define ``highly discussed'' as GFIs where the number of comments (including both issue comments and PRorCommit comments) is one standard deviation (8.35) higher than the mean (9.13) in the GFIs with both expert and newcomer participants. 
Therefore, we filtered out all the GFIs with less than 18 comments in the dataset. Eventually, we obtained 2,607 GFIs. When analyzing the GFI comments, we focus on the dialogues. The comments of a GFI may contain several dialogues. %, as illustrated in Fig.~\ref{fig:dialog}. 
We drew no more GFIs from our sample when we reached data saturation~\cite{fusch2015we}. It means that no additional information or insights are identified. Eventually, we analyzed 43 GFIs, a total of 1,065 comments from 131 newcomers and 98 experts.

\begin{comment}
\begin{figure}[htp]
\centering 
\includegraphics[width = 7cm]{pic/dialog.pdf}
%\vspace{-0.2cm}
\caption{A piece of comments of a GFI}
\label{fig:dialog}
\end{figure}
\end{comment}

We apply thematic analysis~\cite{cruzes2011recommended} to extract the information related to mentoring. (1) We first randomly selected ten GFIs and analyzed their comments, identifying dialogues on how experts mentor newcomers. For each dialog analyzed, we identified the participants involved and what information they want to convey. (2) We then conducted open coding on these dialogues, grouping codes into  (sub)themes that were conceptually similar. This process produced a codebook that reveals different codes and (sub)themes of mentoring. (3) We used this first set of codes to code the remaining GFIs' comments, revealing additional codes and refining the codebook. (4) We used an iterative process until the GFIs' comments no longer revealed new codes not captured in our existing set of themes (theoretical saturation). (5) We reviewed and modified the initial codes and (sub)themes and found opportunities for merging. (6) Then, we defined the final themes aiming to identify the ``essence'' of each theme.
To limit subjectiveness, the second and third authors engaged individually for manual analysis processes, e.g., inductive open coding. The final inter-rater reliability was 0.78 (Cohen’s kappa), which indicates substantial agreement between the labelers~\cite{warrens2015five}. After completing the labeling process, the two labelers discussed their codes, and the discrepancies were discussed with the first author. 

\textcolor{black}{In the second stage, the goal is to evaluate the frequency and difficulty of these topics through an email survey with newcomers and experts.} 
Based on the sub-themes identified, we design ten Likert-scale questions to obtain respondents' views on each code. The wording of the questionnaire is slightly different for newcomers and experts. Specifically, for newcomers, we ask about the difficulty of each challenge and the importance of mentoring strategies, whereas for experts, we ask their views on the frequency of these challenges faced by newcomers and the importance of each mentoring strategy. To obtain the email addresses of newcomers and experts, we focus on the GFIs having both newcomers' and experts' comments. Then, we extract their public email addresses through GitHub API. Eventually, we extracted 4,972 newcomers' emails and 1,288 experts' emails. First, we conducted a pilot survey with 100 newcomers and 100 experts. Then, we sent our questionnaire to another 400 newcomers and 400 experts. We did not send the questionnaire to all email addresses in order to reduce the interference to developers. Finally, we receive 43 responses, including 11 responses from newcomers and 32 responses from experts. Except for 66 emails not delivered successfully, the response rate is 4.6\%.

\textcolor{black}{On the welcome page, we inform participants about the data handling policy and ask for their consent to use their anonymous data only for research. The developers' participation is entirely voluntary and they can drop out at any time. The ethics committee of our university reviewed and approved the procedure and protocol of our survey.}
\subsection{Results}
\begin{table*}
\caption{Discussed Topics During GFI Resolution Process and Attitude of Newcomers and Experts}
%\vspace{-0.2cm}
\centering
\footnotesize
\label{tab:my-table}
\begin{tabular}{@{}ccl@{}}
\toprule
\rowcolor[HTML]{FFFFFF} 
Theme &
  Sub-theme &
  \multicolumn{1}{c}{\cellcolor[HTML]{FFFFFF}Code} \\ \midrule
\rowcolor[HTML]{EFEFEF} 
\multicolumn{1}{l}{\cellcolor[HTML]{FFFFFF}} &
  understanding &
  \begin{tabular}[c]{@{}l@{}}unclear about issue assignment (3,3), unclear about the goal of the issue (3,3), difficulty in\\ reproducing the issue (3,3), unclear about the dependencies/ priority of different issues (3,4)\end{tabular} \\
\rowcolor[HTML]{FFFFFF} 
\multicolumn{1}{l}{\cellcolor[HTML]{FFFFFF}} &
  implementation &
  \begin{tabular}[c]{@{}l@{}}unclear how to start (4,4), challenges of environment setting (3,4), unnecessary change (3,3), \\ minor issues (e.g., code styles, typos, changelog) (2,4)\end{tabular} \\
\rowcolor[HTML]{EFEFEF} 
\multicolumn{1}{l}{\cellcolor[HTML]{FFFFFF}} &
  testing and management &
  \begin{tabular}[c]{@{}l@{}}version/ branches conflicts (3,3), challenges of building/ testing (2,3), unclear about CI (3,3), \\ unclear how to link PR and issue (1,3)\end{tabular} \\
\rowcolor[HTML]{FFFFFF} 
\multicolumn{1}{l}{\multirow{-7}{*}{\cellcolor[HTML]{FFFFFF}newcomer challenges}} &
  communication &
  communication issues (3,3), no response (4,3) \\ \midrule
\rowcolor[HTML]{EFEFEF} 
\cellcolor[HTML]{FFFFFF} &
  community regulations &
  contributing guideline (5,4), instruction of granularity of PR (4,4), instruction of code style (4,3) \\
\rowcolor[HTML]{FFFFFF} 
\cellcolor[HTML]{FFFFFF} &
  task recommendation &
  recommendation of suitable tasks (4,4)\\
\rowcolor[HTML]{EFEFEF} 
\cellcolor[HTML]{FFFFFF} &
  helping getting started &
  \begin{tabular}[c]{@{}l@{}}clarification of the issue (5,5), pointing out the target files/ codes (4,4), providing references/\\ examples (4,4), giving hints (3,4)\end{tabular} \\
\rowcolor[HTML]{FFFFFF} 
\cellcolor[HTML]{FFFFFF} &
\begin{tabular}[c]{@{}c@{}}implementation feedback  \end{tabular} &
  \begin{tabular}[c]{@{}l@{}}inline comments for newcomer's code (4,4), pointing out better solutions/ improvements (5,4), \\ pointing out errors/ limitations/ repercussions (5,5), change of title (3,2)\end{tabular} \\
\rowcolor[HTML]{EFEFEF} 
\cellcolor[HTML]{FFFFFF} &
  \begin{tabular}[c]{@{}c@{}}suggestion on testing and \\ management\end{tabular} &
  instructions for building/ testing (5,4), management of branches (4,3) \\
\rowcolor[HTML]{FFFFFF} 
\multirow{-9}{*}{\cellcolor[HTML]{FFFFFF}expert mentoring} &
  \multicolumn{1}{c}{\cellcolor[HTML]{FFFFFF}attitudes to newcomers} &
  warm welcome (5,5), helpfulness (5,5), encouragement (4,4), compliment and affirmation (4,4)\\
\bottomrule
\multicolumn{3}{l}{The numbers in parentheses indicate survey results. The first/second number is the median of newcomers'/experts' view. For codes of newcomer} \\
\multicolumn{3}{l}{challenges, 1--5 represent ``\textit{very easy, relatively easy, normal, relatively difficult, very difficult}'' (in newcomer questionnaire) or ``\textit{never, seldom, sometimes,}} \\
\multicolumn{3}{l}{\textit{often, always}'' (in expert questionnaire). For codes of expert mentoring, 1--5 represent ``\textit{not important at all, relatively unimportant, normal, relatively}} \\
\multicolumn{3}{l}{\textit{important, very important}''.} 
\end{tabular}
\end{table*}

We illustrate the topics discussed by newcomers and experts during the GFI resolution process, combining the quotations of their comments and survey results. As shown in Table~\ref{tab:my-table}, we extract 32 codes belonging to two themes and ten sub-themes. Compared with the prior research~\cite{balali2018newcomers,steinmacher2015systematic}, our findings are more fine-grained due to the analysis of the historical data. We report the results grouped by the following two themes.

\subsubsection{Newcomer Challenges}
Newcomers face various challenges when solving GFIs, which belong to four sub-themes. 

\textbf{Understanding.} The first challenge is to understand GFI, which relates both technical and non-technical factors. Newcomers are confused about issue assignments, especially in popular communities. For example, a newcomer said, ``\textit{Hey I saw many people wanted to work on this. Can anybody tell me if anybody is presently working on it...}''.\footnote{\url{https://github.com/mlpack/mlpack/issues/2619}} Due to the limited skills or obscure descriptions, newcomers also have difficulties in understanding issue-self. Specifically, we notice that newcomers are unclear about the goal of issues and have difficulties in reproducing issues. Besides, some issues relate to or depend on other issues. For such issues, newcomers are confused about the priorities. This challenge is ranked as ``\textit{often}'' by experts. For example, a newcomer said, ``\textit{So, it appears to me that it's taking way longer than expected to address issue \#1476. Would it be okay for me to implement a fix for this again and try to actually get this merged? Or should we just wait for \#1476?}''.\footnote{\url{https://github.com/neovim/neovim/issues/1474}}

\textbf{Implementation.} The core challenge is implementing patches. First, many newcomers do not know how to start, which is considered ``\textit{relatively difficult}'' by newcomers. This challenge comes from a lack of understanding of projects and limited experience. For example, a newcomer said, ``\textit{I'd like to ask for a little more specific detail about what is needed in this issue and where to look for the messages that need to be modified}''.\footnote{\url{https://github.com/pytorch/ignite/issues/1479}} Second, some projects lack detailed guidance on environment setting, making newcomers feel confused, as a newcomer said, ``\textit{\textcolor{black}{@developer1} I'm working on it, due to difficulty in dependencies I couldn't proceed further}''.\footnote{\url{https://github.com/pytorch/ignite/issues/1315}} Sometimes although the patches can fix the issue, they can not be merged because of: 1) patches may contain unnecessary changes; %, as an expert said, ``\textit{We don't need to create this new style as of now. You can revert this change}''.\footnote{\url{https://github.com/oppia/oppia-android/issues/1949}} 
2) patches may have some minor issues. Interestingly, although the challenge of minor issues is considered ``\textit{relatively easy}'' by newcomers, experts rank its frequency as ``\textit{often}''.

\textbf{Testing and Management.} After implementation, newcomers need to test and submit their patches, which is challenging for newcomers. Many large OSS projects have multiple active branches under development at once, in parallel~\cite{bird2012assessing}. However, working with multiple branches and multiple developers can be tricky, as an expert commented, ``\textit{you should do separate tickets on separate branches so they can be reviewed and altered separately}''.\footnote{\url{https://github.com/kiwix/kiwix-android/issues/2418}} Newcomers also have difficulty in doing testing and building. Some newcomers even do not test their patches, as an expert suggested, ``\textit{it would also be good to include some tests. I think there is a warning test macro in the typ/tests.rs module}''\footnote{\url{https://github.com/gleam-lang/gleam/issues/876}} Moreover, some newcomers are unclear about GitHub, e.g., they do not know how to link PR and issue. Even though newcomers think it is \textit{very easy}, experts still \textit{sometimes} notice that newcomers encounter this challenge.

\textbf{Communication.} This challenge has been extensively discussed in prior research~\cite{balali2018newcomers,tan2019communicate}. We find that besides language barriers, this challenge also comes from the misunderstanding of some basic norms. For example, an expert commented, ``\textit{please do not `Resolve' Comment. The person who has started the conversation should resolve the comment. And you should always reply to comments }''. We notice that no response is one of the top two difficult challenges for newcomers (the other is ``\textit{unclear how to start}''), which can reduce newcomers' willingness to contribute.  

\subsubsection{Expert Mentoring} To help newcomers solve GFIs successfully, experts offer practical guidance on six sub-themes.

\textbf{Community Regulations.} Nothing can be accomplished without norms or standards. Especially for OSS communities, such a free and loose organization, regulations are essential~\cite{bonaccorsi2003open}. However, newcomers often lack awareness of regulations. We find that experts provide rich guidance in this regard, including clarifying contributing guidelines and providing instructions on PR granularity and code style. Among them, clarifying contributing guidelines is considered ``\textit{very important}'' by newcomers.

%For example, an expert said, ``\textit{Before proceeding, I'd recommend reviewing our contributing guidelines and style guide}''.\footnote{\url{https://github.com/Project-Books/book-project/issues/391}}

\textbf{Task Recommendation.} Task recommendation to newcomers is necessary because they often struggle to find appropriate first tasks and the voluntary, self-organizing distribution of decentralized labor intensifies this challenge~\cite{Balali2020Recommending}. This is the original intention of the GFI mechanism. We find that for some GFIs, someone has been already working on it. In this case, experts will recommend other appropriate tasks to newcomers. For example, an expert suggested, ``\textit{\textcolor{black}{@developer2} sounds good, please check some unassigned issues from here...There certainly can be one that you'd like to work on :)}''.\footnote{\url{https://github.com/pytorch/ignite/issues/1315}}

\textbf{Helping Getting Started.} Taking the first step proverbially costs the most. Many newcomers are still confused about GFIs and do not know where to start. To help newcomers take their first steps, experts usually provide various support. The most common support is the clarification of issue, including explaining causes of issue and clarifying goals of bug fix. This support is considered ``\textit{\textit{very important}}'' by both newcomers and experts. For example, an expert explained, ``\textit{the goal is getting rid of a custom reimplementation of strchr() that isn't nearly as optimized...}''.\footnote{\url{https://github.com/neovim/neovim/issues/1474}} 
Sometimes, experts will directly point out the files or codes that need to be modified to help newcomers quickly locate bugs since locating bugs is difficult and expensive, particularly for large-scale systems~\cite{Saha2013Improving}. To assist newcomers to solve issues faster, we notice that some experts give newcomers hints on possible directions or provide examples and references. For example, an expert said, ``\textit{I think we'll need to make a new function similar to is\_simple\_constant for this as is\_simple\_constant doesn't include lists or records}''.\footnote{\url{https://github.com/gleam-lang/gleam/issues/876}}

\textbf{Implementation Feedback.} Experts usually give comments on newcomers' implementation, which is essential to increase code quality and transfer knowledge. GitHub supports reviewers to write inline comments, providing a clear view of thoughts and fine-grained recommendations. We observe that experts have widely adopted this mechanism. The experts' mentoring is also represented in pointing out limitations of newcomers' patches or suggesting improvements. For example, an expert recommended, ``\textit{Let's use options object (\{isExternal\} instead of isExternal). This makes a better API}''.\footnote{\url{https://github.com/serverless/serverless/issues/8497}} The titles of newcomers' patches are sometimes confusing, therefore experts offer advice in this aspect as well. For example, an expert said, ``\textit{\textcolor{black}{@developer3} Feel free to remove [WIP] from the PR title if this is ready for review}''.\footnote{\url{https://github.com/kubernetes/kops/issues/10009}} However, experts consider it  ``\textit{relatively unimportant}''.

\textbf{Suggestion on Testing and Management.} 
As mentioned above, newcomers face challenges of testing and management, so experts give practical guidance on this aspect. This support is considered ``\textit{very important}'' by newcomers because newcomers are often unfamiliar with workflow of projects. For example, an expert showed a newcomer how to conduct testing and said, ``\textit{In general, any PR that changes something that is visible to the user in the CLI should have a copy-paste terminal session in the `testing done' so that we can see what it would look like from the user perspective}.\footnote{\url{https://github.com/confluentinc/ksql/issues/6843}}

\textbf{Attitudes to Newcomers.} Newcomers might shy away from the project if they are introverted, suffer from social anxiety, or get stuck~\cite{sholler2019ten}. Many experts adopt various strategies to make communities friendly and encourage newcomers to contribute without worries. Although this support is non-technical, it is ranked as ``\textit{very/relatively important}'' by both newcomers and experts. For example, an expert encouraged the newcomer not to be afraid of difficulties and pressure, and he said, ``\textit{Sorry for the wall of text, and it's really not as daunting as it may sound, I promise :). Just start hacking around on things and creating a PR and we'll guide you from there}''.\footnote{\url{https://github.com/neovim/neovim/issues/1474}}

\vspace{4pt}
\noindent\begin{tcolorbox}
\noindent\emph{\textbf{Summary for RQ3:}}
%We extract 32 codes about mentoring in the GFI resolution process, which relate to newcomer challenges and expert mentoring. Regarding challenges, we find that newcomers have difficulty in understanding, implementation, testing, etc. To support newcomers, experts offer a warm welcome and provide practical guidance on community regulations, implementation, etc. 
We identify 14 newcomer challenges and 18 expert mentoring contents.
In particular, newcomers have difficulty in understanding, implementing, testing, etc. To support newcomers, experts offer a warm welcome and provide guidance on community regulations, implementation, etc. 
The attitude towards these codes reveals the difficulty of challenges and the importance of supports, as well as the cognitive differences of newcomers and experts. %%not sure what this attitude means in the results: newcomers and experts hold quite different opinions on some codes, e.g., unclear how to link PR and issue
\end{tcolorbox}

\section{RQ4: Relevance}\label{RQ4}
This question aims to understand the relevance of mentoring for newcomers' onboarding. We formulate the following sub-questions: 1) \textbf{Is mentoring related to whether GFIs can be solved by newcomers?} 2) \textbf{Is mentoring related to whether newcomers continue to contribute?}

\subsection{Methodology}
We construct two generalized linear models (GLM) to answer the two sub-questions. For the first model, we propose two categories of metrics as independent variables considering the factors that may affect newcomers to solve GFIs, especially mentoring behavior.

1) \textit{Issue-Related.} We consider the following four groups of metrics that are directly related to GFIs. 

$\bullet$ \textit{Basic Characteristics.} If an issue is opened for a long time, it may indicate that the issue is difficult and is not suitable for newcomers. Therefore, we consider ``\textbf{duration}'' as one metric. It is defined as the interval between an issue's open time and close time. We also choose the number of labels of a GFI --- ``\textbf{\#label }'' as a variable. We hypothesize that it may be easier for newcomers to solve if a GFI has more labels.

$\bullet$ \textit{Expert-Related Factors.} First, we treat whether experts are involved in the resolution process of GFI --- ``\textbf{expert\_involvement}'' as one categorical variable because we assume if a GFI has expert participation, it may be easier for newcomers to solve. Similarly, we also include the number of experts involved in --- ``\textbf{\#expert }'' as another metric. Second, we hypothesize that if experts leave more comments on the GFI, newcomers will be more likely to solve this GFI. Therefore, we include the number of expert comments --- ``\textbf{\#expert\_comment}'' as the third metric.

$\bullet$ \textit{Newcomer-Related Factors.} First, we hypothesize that if a GFI has more newcomer participants, it may be more likely to be solved by newcomers. Therefore, we consider the number of newcomers --- ``\textbf{\#newcomer}'' as one metric. Second, we hypothesize that more comments left by newcomers may suggest that newcomers are more willing and spend more effort on the GFI. Therefore, we consider the number of newcomer comments --- ``\textbf{\#newcomer\_comment}'' as another metric. 

$\bullet$ \textit{Solution-Related Factors.} The solution of the GFI can reflect its’ difficulty, so we consider these kinds of factors. The following metrics measure the commits corresponding to a GFI and are borrowed from a study evaluating software changes~\cite{Shihab2012Industrial}. The first one is the number of modified files --- ``\textbf{\#file }'' because we assume that if more files were modified, the GFI may be more difficult~\cite{Eick2001Does}. Similarly, the second one is the number of modified code lines --- ``\textbf{\#line}''.  

2) \textit{Project-Related.} We consider two metrics that are directly related to projects. First, if a project is more popular, it will make the GFI more likely to attract newcomers’ attention~\cite{qiu2019signals}. Therefore, we treat the number of stars --- ``\textbf{\#star}'' as a variable. Second, we find four types of mentorship structures of OSS communities in RQ2. We assume that mentorship structures may be related to the resolution of GFI, so we consider ``\textbf{mentorship\_network}'' as a categorical variable.

We also include the age of the project --- ``\textbf{\#month}'', the number of authors --- ``\textbf{\#author}'', and the number of commits --- ``\textbf{\#commit}'' as control variables.

For the second sub-question, we also consider the above metrics. However, the based hypotheses are slightly different. For example, for \textbf{\#newcomer\_comment}, we hypothesize that if newcomers leave more comments, it suggests that they are more willing to join the community and are more likely to make further contributions. 

The dependent variable Y of the first model is whether a GFI is solved by a newcomer. We code the dependent variable as Y = 1 if a GFI is solved by a newcomer and Y = 0 otherwise. For the second model, the dependent variable Y is whether a newcomer who solved GFI makes further contributions. We code the dependent variable as Y = 1 if a newcomer makes further contributions and Y = 0 otherwise.

To fit the models, we need to know who solved a GFI. However, there is no such field in GitHub and manual analysis is impossible for 48,402 GFIs. Referring to the method of Tan et al.~\cite{tan2020first}, we focus on the GFIs that were automatically closed by the corresponding commits. Eventually, we obtain 3,875 GFIs using this mechanism, including 1,020 cases solved by newcomers and 2,855 cases solved by experienced developers. To fit the first model, we use all the 1,020 GFIs solved by newcomers and randomly select 1,020 GFIs solved by experienced developers. To fit the second model, we only focus on the 1,020 GFIs solved by experienced developers, including 496 GFIs solved by one-time contributors and 524 GFIs solved by newcomers who would continue to contribute later.

Then, we calculate the Spearman rank correlation for the metrics to test whether there are any correlated pairs of metrics. We find that for the two models, four pairs of metrics have high correlations, i.e., correlation coefficient $d \geqslant 0.7$~\cite{tan2020first, mcintosh2016empirical}: \textbf{\#expert} and \textbf{\#expert\_comment}, \textbf{\#newcomer} and \textbf{\#newcomer\_comment}, \textbf{\#author} and \textbf{\#star}, \textbf{\#commit} and \textbf{\#star}.
We decide to keep the latter. For the second model, we also find that \textbf{\#file} and \textbf{\#line} are highly correlated, for which we decide to keep \textbf{\#line}. We use GLM provided by the R package and adopt the stepwise regression method to find the final models.

\subsection{Results}
\subsubsection{Resolution of GFIs}
Table~\ref{tab:RQ4-1} shows the results of the first model. We use McFadden's Pseudo R-Squared~\cite{walker2016nine} to evaluate goodness-of-fit, which yields a $R^2$ of 0.29 ($> 0.2$), representing an excellent fit~\cite{Zhou2018What}. We find five significant predictors. The most significant factors are \textbf{\#newcomer\_comment} and \textbf{\#star}. The results show that if a GFI has more newcomer comments and exists in a popular project, it has more chance to be solved by newcomers. This is reasonable because more comments mean more effort, and naturally, the newcomer is more likely to solve the GFI. The popularity of a project is also important, if a project is less popular, it is more difficult to attract newcomers' attention. The results also show that if GFI needs a longer time (\textbf{duration}) to be closed, it may have less chance to be solved by newcomers. The longer the time, the more difficult the GFI may be. 
The predictors --- \textbf{expert\_involvement} and \textbf{\#expert\_comment} are the most concerned factors. We observe that \textbf{expert\_involvement} is positively correlated with GFI being solved by newcomers. It suggests that expert involvement is important for newcomers' successful contribution and just recommending tasks to newcomers may not be enough. However, \textbf{\#expert\_comment} shows a negative relationship. After all, the more the experts' comments, the more difficult the GFI may be. We do not observe the effects of different mentorship structures.
\begin{table}[ht]
\footnotesize
\centering
\caption{\textcolor{black}{Results of the GLM for Resolution of GFIs. \\Only Significant Predictors are Shown} }
%\vspace{-0.2cm}
\label{tab:RQ4-1}
\begin{tabular}{@{}lcccc@{}}
\toprule
\multicolumn{1}{c}{Metrics} & \multicolumn{1}{c}{Est.} & \multicolumn{1}{c}{Std. Err.} & \multicolumn{1}{c}{z} & \multicolumn{1}{c}{$Pr(>|z|)$} \\ \midrule
(Intercept)          & -2.85 & 0.20 & -14.44 & \textless 2e-16 \\
\#expert\_comment    & -0.41 & 0.13 & -3.20  & 0.00138         \\
\#newcomer\_comment  & 1.57  & 0.11 & 14.79  & \textless 2e-16 \\
duration             & -0.41 & 0.07 & -6.03  & 1.6e-9          \\
\#star               & 0.30  & 0.03 & 11.09  & \textless 2e-16 \\
expert\_involvementY & 0.71  & 0.19 & 3.83   & 0.00013         \\ \bottomrule
\end{tabular}
\leftline{\footnotesize{\qquad All the non-categorical variables were log-transformed.}}
\end{table}

\begin{table}[]
\footnotesize
\centering
\caption{\textcolor{black}{Results of the GLM for Resolution of GFIs. \\Only Significant Predictors are Shown}}
%\vspace{-0.2cm}
\label{tab:RQ4-2}
\begin{tabular}{@{}lcccc@{}}
\toprule
\multicolumn{1}{c}{Metrics} & \multicolumn{1}{c}{Est.} & \multicolumn{1}{c}{Std. Err.} & \multicolumn{1}{c}{z} & \multicolumn{1}{c}{$Pr(>|z|)$} \\ \midrule
(Intercept)          & 1.64  & 0.28 & 5.86  & 4.73e-09 \\
\#newcomer\_comment  & 0.20  & 0.09 & 2.34  & 0.0192   \\
\#star               & -0.17 & 0.03 & -5.29 & 1.23e-07 \\
\#file               & 0.17  & 0.08 & 2.17  & 0.0304   \\
expert\_involvementY & -0.74 & 0.16 & -4.52 & 6.24e-06 \\ 
\bottomrule
\end{tabular}
\leftline{\footnotesize{\qquad All the non-categorical variables were log-transformed.}}
\end{table}

\subsubsection{Retention of Newcomers}
Table~\ref{tab:RQ4-2} shows the results of the second model. The McFadden's Pseudo R-Squared~\cite{walker2016nine} is 0.05 ($<0.2$), indicating not a perfect fit. It means that the factors affecting the newcomers' retention are complex, so it can not be well explained just based on newcomers' initial contributions. However, the model is still acceptable because the model objective is to show the importance of the particular predictors on the response instead of prediction~\cite{grace2012can}. There are four significant predictors. We find that if newcomers leave more comments on a GFI (\textbf{\#newcomer\_comment}), the GFI may have more chances to be solved by newcomers. It reflects that newcomers' willingness is essential for their further contribution, consistent with the findings of Zhou et al~\cite{zhou2012make}. We also observe that if a GFI involves more files modified (\textbf{\#file}), newcomers are more likely to solve it. It indicates that newcomers' previous skill is important for their retention. Surprisingly, the results of other predictors are different from our hypotheses. We find that if a newcomer successfully solves the GFI in a less popular project (\textbf{\#star}), she/he is more likely to make further contributions. It may indicate that newcomers are more willing to contribute so they are more likely to retain. Moreover, we find that \textbf{expert\_involvement} is negatively correlated with newcomer retention. It means that the newcomers who successfully solved GFIs without expert involvement are more likely to make further contributions. This result also strengthens the above findings, i.e., newcomers' previous skills and willingness are important factors for their retention.

\vspace{4pt}
\noindent\begin{tcolorbox}
\noindent\emph{\textbf{Summary for RQ4:}}
We find that newcomers will have more chances to solve the GFI if: 1) they leave more comments on the GFI; 2) the GFI is from more popular projects; and 3) experts are involved in the resolution process.
%spend more effort on GFIs reported in more popular projects %%unclear sentence, re-phrase it
%and experts are involved in the resolution process, they may have more chances to solve the GFIs. 
However, the factors related to newcomers' retention are complex and can not be well explained only by their initial contributions. Nevertheless, newcomers with stronger skills and stronger willingness are more likely to make further contributions.
\end{tcolorbox}

\section{Discussion and Implication}\label{implications}
We discuss the significance of mentoring for newcomers' onboarding based on our findings and point out implications for practitioners and researchers.
\subsection{Significance of Mentoring}
Our findings show that 70\% of GFIs have expert involvement, which is positively correlated with the probability of newcomers' successful contribution. However, we find that expert involvement is negatively correlated with newcomers' retention. On the surface, the two opposite correlations make mentoring tricky. 

We argue that in fact, mentoring is essential for newcomers’ onboarding. First, expert mentoring can mitigate the challenges that newcomers face during their initial contributions. For example, RQ3 finds that newcomers are usually unclear about how to start. To reduce this barrier, experts provide various practical supports, e.g., clarification of the issue, pointing out the target files/codes, and giving hints. Second, the perfect-fitting model for resolution of GFIs shows that expert involvement is a significant predictor (see Table~\ref{tab:RQ4-1}). The Est. value indicates a 71\% increase in newcomers’ initial contributing success with expert involvement compared to the GFIs without expert involvement. Third, although the regression model about retention of newcomers shows that expert involvement is a negative predictor, it should be noticed that the retention can not be well explained just by newcomers' initial contribution. Actually, among the newcomers who make further contributions, 64.4\% of them have expert participation during the GFI resolution process. Combining with other significant predictors, we presume that the phenomenon that newcomers can successfully solve GFIs without expert involvement is more likely to essentially reflect their strong ability and willingness. Future studies need to investigate how to directly measure the ability and willingness of newcomers, thus reducing the interference to the factor of expert involvement.

\subsection{Implication}
We conduct the first detailed analysis of historical data on daily mentoring behavior for newcomers' onboarding in OSS communities, which has rich implications for practitioners and researchers.

The comprehensive illustration of mentoring behavior and its importance can promote mentoring process and in turn, reduce the barriers to newcomers' onboarding. In particular, RQ3 identifies the challenges of newcomers during their initial contributions. Experts can plan ahead and provide more targeted guidance on these challenges. For example, we find that ``no response'' is a common challenge faced by newcomers. Even for the GFIs with experts' responses, the response time for a few of them is more than two months (RQ1). To mitigate this challenge, experts can adopt some automated reminder bots to remind to answer newcomers' questions. We also observe that newcomers and experts have different views on certain codes, e.g., minor issues. Therefore, revealing such cognitive differences can make mentoring process more efficient. Through RQ2, we find four types of mentorship structures in OSS communities, which reveals the social relationship between experts and newcomers. Among the four types, it is worth noting that the communities of ``centralized mentoring'' may have a higher risk of single-point failure if the sole expert leaves or becomes inactive. Therefore, the communities need to adopt some incentive mechanisms to encourage and cultivate backup mentors. Moreover, RQ4 reveals various factors related to the success of newcomers' initial contributions and their retention, which can motivate various practical strategies supporting newcomers' onboarding. For example, we find that \#newcomer\_comment is a positive factor. Therefore, experts should provide a friendly and inclusive atmosphere to encourage newcomers to express their opinions as much as possible.   

More research will be necessary to refine and further elaborate our findings, which provide a basis for researchers to ask other research questions. For example, although we try to analyze the relationship between mentorship structures and newcomers' onboarding, this predictor is not significant may due to the limited labeled data. Future research can design classification algorithms based on the different mentorship structures we identified, so as to expand the labeled data. Future research can also introduce other factors to refine our models, especially the model of newcomers' retention. Designing automatic tools to facilitate mentoring process based on the challenges identified in RQ3 is also a significant direction.

\section{Threats to Validity}\label{threats}
\textbf{External Validity} relates to the generalizability of our findings. When conducting this study, we only focus on the top 1\% repositories with most GFIs on GitHub. Although we believe onboarding newcomers is important for all projects, the mentoring behavior is rare and hard to identify in the communities with few GFIs. However, we believe that all communities may benefit from our findings, especially the communities that neglect to provide guidance to newcomers.

\textbf{Internal Validity} concerns the threats to how we perform our
study. The first threat comes from the way of identifying mentoring behavior. In this study, we treat if a GFI has experts' comments, there exists mentoring behavior. This may bring risks in some cases, e.g., the GFIs do not attract newcomers' attention, and the experts' comments do not directly answer newcomers' questions. Nevertheless, we still believe that since GFI is specifically for newcomers, any comments from experts may provide effective mentoring for (potential) newcomers to solve the issue. The second threat comes from our qualitative analysis for RQ2\&3. Although we select a sample of projects to explore the mentorship structures (RQ2), the margins of error are acceptable~\cite{kotrlik2001organizational}. As for RQ3, in order to identify the discussed topics, we follow the advice of Fusch et al.~\cite{fusch2015we} to introduce new GFIs until we reach data saturation. For the manual analysis in RQ3, we acknowledge that the choice of some codes is to some extent subjective. To mitigate this risk, the initial coding is performed independently by the two authors of this paper. After this initial proposal, we compare our list of codes. Then, based on a coding guide or performing peer-review on each result, the data finally received the full agreement. The last threat comes from the regression models for RQ4. Through RQ4, we aim to understand the importance of mentoring for newcomers' onboarding. We acknowledge that the correlations between variables reveal that there exist certain patterns in the data, which cannot prove causal relationships. Therefore, future studies can try to explain the causal relationship between mentoring and newcomers' onboarding.

\textbf{Construct Validity} concerns the relationship between the treatment and the outcome. The threat comes from the rationality of the questions asked. We are interested in understanding daily mentoring behavior in OSS projects. To achieve this goal we focus on the resolution process of GFIs, and analyze the involvement of experts, mentorship structures, discussed topics, and the relevance of experts' participation. We believe these questions can help us better understand mentoring behavior and support newcomers' onboarding.

\section{Conclusion}\label{conclusion} 
This paper presents the first large-scale empirical study of daily mentoring in OSS communities. Based on the historical data of 48,402 GFIs from 964 repositories, we investigate the extent of expert involvement during GFIs' resolution process, the mentorship structures and their properties towards the community, the discussed topics related to mentoring, and the relevance of mentoring to the success of newcomers' first contribution and newcomers' retention. Our study reveals a number of interesting findings and illustrates the significance of mentoring for newcomers' onboarding. Moreover, we discuss the implications and recommendations for practitioners and provoke different research directions for future OSS researchers. Newcomers' onboarding is a critical issue for OSS communities. To reduce the barriers, many OSS communities and researchers spend a lot of effort recommending suitable tasks for newcomers. However, they are unclear or even neglect to mentor newcomers in this process. Our study investigates mentoring from multiple perspectives, which can help us better understand the significance of mentoring in OSS communities. 

%%\vspace{-0.2cm}
\section{Data Availability}
To facilitate replications or other types of future work, we provide the data and scripts online: \url{https://figshare.com/s/addd697d581c82f96f9a}.

\section{Acknowledgements}
This work is supported by the National Natural Science Foundation of China Grants 62202022, 62141209, 61825201, and Self-determined Research Funds of State Key Laboratory of Software Development Environment SKLSDE-2022ZX-08. 

\balance
\bibliographystyle{IEEEtran}
\bibliography{reference}
\end{document}